\documentclass[12pt]{article}
\pdfoutput=1
\usepackage{pstricks}
\usepackage{color}
\usepackage{amssymb,amsmath,bm,bbm}
\usepackage{epsf}
\usepackage{epsfig}
\usepackage{afterpage}
\usepackage{longtable}
\usepackage{cite}
\usepackage{latexsym,mathrsfs,dsfont}
\usepackage{graphics}
\usepackage{url}
\usepackage{paralist}
\usepackage{bbold}

\newcommand{\IM}{\rm{Im}}
\newcommand{\RE}{\rm{Re}}

\newcommand{\tev}{\, {\rm TeV}}
\newcommand{\gev}{\, {\rm GeV}}
\newcommand{\mev}{\, {\rm MeV}}

\newcommand{\vcb}{|V_{cb}|}

\newcommand{\vub}{|V_{ub}|}

\newcommand{\bsi}{B_6^{(1/2)}}
\newcommand{\bei}{B_8^{(3/2)}}

\def\epe{\varepsilon'/\varepsilon}
\newcommand{\beq}{\begin{equation}}
\newcommand{\eeq}{\end{equation}}
\newcommand{\be}{\begin{equation}}
\newcommand{\ee}{\end{equation}}
\newcommand{\bi}{\begin{itemize}}
\newcommand{\ei}{\end{itemize}}
\newcommand{\ba}{\begin{array}}
\newcommand{\ea}{\end{array}}
\newcommand{\beqa}{\begin{eqnarray}}
\newcommand{\eeqa}{\end{eqnarray}}
\newcommand{\bea}{\begin{eqnarray}}
\newcommand{\eea}{\end{eqnarray}}
\newcommand{\beqn}{\begin{eqnarray}}
\newcommand{\eeqn}{\end{eqnarray}}

\newcommand{\D}{\Delta}

\newcommand{\eps}{\epsilon}

\definecolor{red}{cmyk}{0,1,1,0.4}

\def\kpn{K^+\rightarrow\pi^+\nu\bar\nu}
\def\klpn{K_{L}\rightarrow\pi^0\nu\bar\nu}

\usepackage{fancyhdr}
\advance \headheight by 3.0truept       

\begin{document}

\begin{flushright}
    {FLAVOUR(267104)-ERC-114}\\
    {BARI-TH/15-700}
\end{flushright}

\medskip

\begin{center}
{\LARGE\bf
\boldmath{$\epe$ in 331 Models}}
\\[0.8 cm]
{\large\bf Andrzej~J.~Buras$^{a,b}$ and Fulvia~De~Fazio$^{c}$ 
 \\[0.5 cm]}
{\small
$^a$TUM Institute for Advanced Study, Lichtenbergstr. 2a, D-85747 Garching, Germany\\
$^b$Physik Department, Technische Universit\"at M\"unchen,
James-Franck-Stra{\ss}e, \\D-85747 Garching, Germany\\
$^c$Istituto Nazionale di Fisica Nucleare, Sezione di Bari, Via Orabona 4,
I-70126 Bari, Italy}
\end{center}

\vskip0.41cm


\abstract{%
\noindent
Motivated by the recent findings that the ratio $\epe$ in the Standard Model (SM)  appears to be significantly below
the data we investigate whether the necessary  enhancement of $\epe$
can be obtained in
331 models, based on the gauge group $SU(3)_C\times SU(3)_L\times U(1)_X$. 
In these models new physics (NP) contributions to $\epe$ and other flavour observables are  dominated by tree-level exchanges of a $Z^\prime$ with non-negligible contributions from tree-level $Z$ 
exchanges generated through the $Z-Z^\prime$ mixing.  NP contributions to $\epe$ in these models are governed by the electroweak penguin operator $Q_8$ for which the hadronic matrix element is already rather well known so that our analysis of NP contributions
is subject to much smaller theoretical uncertainties than within the SM. In particular 
strong cancellations between different contributions do not take place.
The size of NP effects
depends not only on $M_{Z^\prime}$ but in particular on a parameter $\beta$, which distinguishes between various 331 models. Also a parameter $\tan\bar\beta$ 
present in the $Z-Z^\prime$ mixing plays a role. We perform the 
analysis in seven 331 models characterized by different $\beta$, $\tan\bar\beta$  and fermion representations under the gauge group that have been selected 
in our earlier analysis on the basis of electroweak precision data. Imposing the constraints from $\Delta F=2$ transitions we find that only three of these 
models can provide a significant positive shift in $\epe$ up to $6\times 10^{-4}$ for
 $M_{Z^\prime}=3\tev$. Two of them allow simultaneously a supression of the 
rate for   $B_{s}\to \mu^+\mu^-$ by $20\%$ thereby bringing the  theory closer to the data without any significant impact 
on the Wilson coefficient $C_9$.  The third model provides simultaneous 
shift  $\Delta C_9=-0.6$, softening the anomalies in $B\to K^*\mu^+\mu^-$, without any significant impact on  $B_{s}\to \mu^+\mu^-$.
NP effects in rare $K$ decays, in particular in $\kpn$, turn out to be 
 small. This is also the case of $B\to K(K^*)\nu\bar\nu$. Both predictions 
could be challenged by NA62 experiment and Belle II in this decade.
 The special flavour structure of 331 models implies that  even for  $M_{Z^\prime}=30\tev$ a shift of $\epe$ up to $8\times 10^{-4}$ and a significant shift in 
$\varepsilon_K$ can be obtained, while the effects in other flavour observables
are small. In this manner $\epe$ and $\varepsilon_K$ appear to be  unique flavour observables in these models which provide the possibility of accessing masses of $M_{Z^\prime}$ far beyond the LHC reach.

\thispagestyle{empty}
\newpage
\setcounter{page}{1}

\tableofcontents

\section{Introduction}
The recent findings that the ratio $\epe$  predicted by the Standard Model (SM) appears to be significantly below 
the  experimental data  \cite{Buras:2015yba} poses a natural question what kind 
of new physics (NP) could be responsible for this new anomaly. In the present 
paper we will address this question in  331 models based on the gauge group $SU(3)_C\times SU(3)_L\times U(1)_X$   \cite{Pisano:1991ee,Frampton:1992wt}. 
 These models display several appealing features. 
The first one is that the requirement of asymptotic freedom of QCD  together with that of  anomaly cancelation constrains the number of generations to be necessarily equal to the number of colours,  providing an explanation for the existence of three generations.
Moreover,  under the action of $SU(3)_L$  two quark generations  should transform as triplets, one as an antitriplet.  
Adopting the choice that the third generation is the one transforming as an antitriplet, 
 this different treatment could be at the origin of the large top mass.
It should be recalled that some of the generators of the group are connected by the relation $Q=T_3+\beta T_8+X$ where $Q$ is the electric charge, $T_3$ and $T_8$ are two of the $SU(3)$ generators and $X$  the generator of $U(1)_X$. $\beta$ is a  parameter that defines a specific variant of the model. 

Several new particles are present in these models, their features depending on the chosen variant.  However,  in all the variants a new neutral gauge boson $Z^\prime$ exists that can mediate tree level flavour changing neutral currents (FCNC) in the quark sector. 

In the framework of 331 models, the ratio  $\epe$ has been studied in 
 our earlier analysis  \cite{Buras:2014yna}. Here we update it and improve  it. 
 
 Recent analyses 
 addressing the implications of new value of $\epe$ within the SM for NP in the
Littlest Higgs model with T-parity and simplified $Z$ and $Z^\prime$ models 
can be found in \cite{Blanke:2015wba} and \cite{Buras:2015yca}, respectively.

The present status of $\epe$ in the SM has been reviewed recently in \cite{Buras:2015yba}, where references to rich literature can be found.
 After the new results for the hadronic matrix elements of QCD penguin and 
electroweak penguin $(V-A)\otimes (V+A)$ operators from  RBC-UKQCD lattice
collaboration \cite{Blum:2015ywa,Bai:2015nea} and the extraction of the
corresponding matrix elements of penguin  $(V-A)\otimes (V-A)$ operators  from the CP-conserving 
$K\to\pi\pi$ amplitudes one finds \cite{Buras:2015yba}
\be\label{LBGJJ}
   \epe = (1.9 \pm 4.5) \times 10^{-4} \,.
\ee
This result differs with $2.9\,\sigma$ significance from 
the experimental world average
from NA48 \cite{Batley:2002gn} and KTeV
\cite{AlaviHarati:2002ye,Abouzaid:2010ny} collaborations, 
\be\label{EXP}
(\epe)_\text{exp}=(16.6\pm 2.3)\times 10^{-4} \,,
\ee
suggesting evidence for NP in $K$ decays. 

But even discarding the lattice results and using instead newly derived upper 
bounds on the matrix elements of the dominant penguin operators from large $N$ 
approach \cite{Buras:2015xba}, one finds at most \cite{Buras:2015yba} 
\be\label{BoundBGJJ}
(\epe)_\text{SM}= (8.6\pm 3.2) \times 10^{-4} \,,
\ee
still $2\,\sigma$  below the experimental data.

The dominant uncertainty in the SM prediction for $\epe$ originates from 
the partial cancellation between QCD penguin contributions and electroweak 
penguin contributions that depend sensitively on the parameters $\bsi$ and 
$\bei$, respectively. QCD penguins give a positive contribution while electroweak penguins
a negative one. Fortunately a new insight in the values of these parameters 
has been obtained recently through the results from the RBC-UKQCD collaboration on 
the relevant hadronic matrix elements of the operators $Q_6$ \cite{Bai:2015nea}
and $Q_8$ \cite{Blum:2015ywa} and  upper bounds on both $\bsi$ and $\bei$ which 
can be derived from large $N$ approach \cite{Buras:2015xba}.

 The  results from  the RBC-UKQCD collaboration  imply the following values 
for $\bsi$ and $\bei$ \cite{Buras:2015yba,Buras:2015qea}
\be\label{Lbsi}
\bsi=0.57\pm 0.19\,, \qquad \bei= 0.76\pm 0.05\,, \qquad (\mbox{RBC-UKQCD})
\ee
 and the bounds from large $N$ approach 
 read
\cite{Buras:2015xba}
\be\label{NBOUND}
\bsi\le \bei < 1 \, \qquad (\mbox{\rm large-}N).
\ee
While one finds in this approach $B_8^{(3/2)}(m_c)=0.80\pm 0.10$, the result for $\bsi$ is less
precise but there is a strong indication that $\bsi < \bei$ in agreement with
(\ref{Lbsi}). For further details, see \cite{Buras:2015xba}.

 In 331 models we have
\be\label{total}
\left(\frac{\varepsilon'}{\varepsilon}\right)_{331}=\left(\frac{\varepsilon'}{\varepsilon}\right)_{\rm SM}+\left(\frac{\varepsilon'}{\varepsilon}\right)_{Z^\prime}+
\left(\frac{\varepsilon'}{\varepsilon}\right)_{Z}\,\equiv 
\left(\frac{\varepsilon'}{\varepsilon}\right)_{\rm SM}+\Delta(\epe)
\ee
with the $\Delta(\epe)$  resulting from tree-level $Z^\prime$ and $Z$ exchanges.
In this paper we will concentrate exclusively on this shift in $\epe$, which 
as we will see has significantly smaller theoretical uncertainties than the 
SM part.

Indeed, as demonstrated by us in  \cite{Buras:2014yna}, the shift in $\epe$ in question
is in 331 models governed by the electroweak $(V-A)\times (V+A)$ penguin operator 
\begin{equation}\label{O4} 
Q_8 = \frac{3}{2}\,(\bar s_{\alpha} d_{\beta})_{V-A}\!\!\sum_{q=u,d,s,c,b,t}
      e_q\,(\bar q_{\beta} q_{\alpha})_{V+A} \,
\end{equation}
with only small contributions of other operators that we will neglect in what 
follows.

As the relevant non-perturbative parameter $\bei$ is much better known 
than $\bsi$, the non-perturbative uncertainty in NP contributions in (\ref{total}) is significantly smaller than in the SM term. Moreover, except for 
the value of $\bei$, the NP contributions are fully independent from the 
SM one. Consequently we can fully concentrate on the these new contributions 
and investigate which 331 models can bring theory closer to data. It will also be interesting to see what this implies for other flavour observables, 
in particular branching ratios for $B_{s,d}\to\mu^+\mu^-$, $B\to K(K^*)\nu\bar\nu$, $\kpn$, $\klpn$ and the Wilson coefficient $C_9$ that enters the discussion of $B\to K^*\mu^+\mu^-$ anomalies. 

In this context
our detailed analyses of FCNC processes
in 331 models in \cite{Buras:2012dp,Buras:2013dea} will turn out to be useful. 
References to earlier analyses of flavour physics in 331 models 
can be found there and in  \cite{Diaz:2004fs,CarcamoHernandez:2005ka}. But the ratio $\epe$ has been analyzed in 331 models only in  
\cite{Buras:2014yna}. However, in that paper  values of $\bsi$ as high as 1.25 and thus violating 
the bounds in (\ref{NBOUND}) have been considered. As seen in Fig.~14 of 
that paper  in this case  the resulting $\epe$ in the SM can even be 
larger than the data so that dependently on the chosen value of $\bsi$ 
both enhancements and suppressions of $\epe$ in a given model were required to 
fit the data in (\ref{EXP}) with different implications for $\klpn$. With 
the new results in (\ref{Lbsi}) and (\ref{NBOUND}) the situation changed 
drastically and one needs a significant enhancement relative to the SM prediction.

The main new aspects of the present paper relative to \cite{Buras:2014yna} are as follows:
\begin{itemize}
\item
We update our analysis of $\Delta(\epe)$ by taking new result on $\bei$ 
into  account. We also include in the discussion second fermion representation 
($F_2$) which was not considered by us in the context of $\epe$ previously. 
\item
After the constraints from $\Delta F=2$ transitions have been taken into 
account the size of the possible enhancement of $\epe$ depends on a given model and in certain models it
is too small to be relevant. Such models are then disfavoured.
\item
Further selection of the models is provided through the correlation of 
$\epe$ with other flavour observables, in particular the decays $B_{s}\to\mu^+\mu^-$, 
$B\to K^*\mu^+\mu^-$, $\klpn$, $\kpn$ and 
$B\to K(K^*)\nu\bar\nu$. While a definite selection 
is not possible at present, because the data is not sufficiently precise,
it will be possible in the coming years.
\end{itemize}

In  \cite{Buras:2014yna} we have considered several  331 models corresponding to four different values 
of $\beta$, three values of  $\tan\bar\beta$ related to $Z-Z^\prime$ mixing and 
two fermion representations $F_1$ and $F_2$. 24 models in total. Among them 7 have been favoured by electroweak precision tests and we will concentrate our 
analysis on them.
The important result of the present paper is 
that the requirement of significant enhancement  of $\epe$ reduces the number 
of favourite  models to 3.

Our paper is organized as follows. In Section~\ref{sec:2} we present 
the general formula for $\epe$ in 331 models which is now  valid for both 
fermion representations considered in \cite{Buras:2014yna}. In Section~\ref{sec:3} we first briefly introduce the 7 favourite 331 
models that have been selected in  \cite{Buras:2014yna} on the basis of 
electroweak precision data. Subsequently, imposing the constraints from 
$\Delta F=2$ transitions in $K$ and $B_{s,d}$ systems, we find that only 
three models are of interest for $\epe$. Subsequently we demonstrate that
 the correlations of $\Delta(\epe)$ with rare decays in  these three
 models can provide further selection
 between them when the data on flavour observables considered 
by us improves. In Section~\ref{sec:4} we consider the case of a $Z^\prime$ 
outside the reach of the LHC. The particular flavour structure of 331 models 
implies that for $M_{Z^\prime}\ge 10\tev$, NP effects in rare $B_{s,d}$ decays, 
$\kpn$ and $\klpn$ are very small, while the ones in $\epsilon_K$ and $\epe$ can be significant  even for  $M_{Z^\prime}= 30\tev$.
 We conclude in Section~\ref{sec:5}.
Except for the formulae for $\epe$, all other expressions for observables 
considered by us can be found in 
\cite{Buras:2014yna,Buras:2012dp,Buras:2013dea} and we will not repeat them 
here. But  in Table~\ref{tab:input} we give all relevant input 
parameters which occassionaly differ from the ones used in 
\cite{Buras:2014yna,Buras:2012dp,Buras:2013dea}.

\boldmath
\section{Basic Formula for $\epe$ in 331 Models}\label{sec:2}
\unboldmath
\subsection{Preliminaries}
In 331 models $\epe$ receives the dominant new contribution from tree-level $Z^\prime$ exchanges but through $Z-Z^\prime$ mixing, analyzed in detail in \cite{Buras:2014yna}, contributions from tree-level $Z$ exchanges in certain models and for certain values of new parameters cannot be neglected. We begin with general expressions 
valid for both $Z^\prime$ and $Z$ contributions which will allow us to recall all the relevant parameters of the 331 models.  Subsequently we will 
specify these expressions to $Z^\prime$ and $Z$ cases which differ only by the value of the Wilson coefficient of the $Q_8$ operator at the 
low renormalization scale at which the relevant hadronic matrix element of 
$Q_8$ is calculated.

The basic expression for $V=(Z^\prime,Z)$ contribution  to $\epe$ is given by
\be\label{eprimeZfinal}
\left(\frac{\varepsilon'}{\varepsilon}\right)_{V}=
\frac{\omega}{|\varepsilon_K|\sqrt{2}}\frac{{\IM} A_2(V)}{{\RE}A_2}, \qquad
\omega=\frac{{\rm Re}A_2}{{\rm Re}A_0}= (4.46)\times 10^{-2}\,.
\ee
The factor $\omega$ differs  by $10\%$ from the corresponding factor $\omega_+$ in 
\cite{Buras:2014yna} as discussed in \cite{Buras:2015yba}.

In evaluating (\ref{eprimeZfinal}) we use, as in the case of the SM, the experimental values for  ${\rm Re} A_2$ and $\varepsilon_K$:
\be\label{A0A2}
{\rm Re}A_2= 1.210(2)   \times 10^{-8}~\gev,\qquad
|\varepsilon_K|=2.228(11)\times 10^{-3}\,\,.
\ee

The amplitude  $A_2(V)$ is dominated by the contribution of the $Q_8$ 
operator 
and is given by 
\be\label{RENPZ2}
A_2(V)= C_8(m_c,V)\langle Q_8 (m_c)\rangle_2\,
\ee
with $C_8(m_c,V)$ given below.

The hadronic matrix element is given by 
\be
\label{eq:Q82}
\langle Q_8(m_c) \rangle_2 = \sqrt{2} 
\left[ \frac{m_{\rm K}^2}{m_s(m_c) + m_d(m_c)}\right]^2 F_\pi \,B_8^{(3/2)}=0.862\,\bei\,\gev^3.
\ee
The normalization of our amplitudes is such that $\langle Q_8(m_c) \rangle_2 $
is by a factor of $\sqrt{\frac{3}{2}}$ smaller than 
the one in \cite{Buras:2015yba}. Correspondingly the value of ${\rm Re} A_2$ in (\ref{A0A2}) is by this factor smaller 
than in  \cite{Buras:2015yba}.
The choice of the scale $\mu=m_c$ is convenient as it is used in analytic 
formulae for $\epe$ in \cite{Buras:2015yba}. 
In our numerical calculations we will use $\bei=0.76$ and the values  
\cite{Agashe:2014kda,Aoki:2013ldr}
\be\label{FpFK}
m_K=497.614\mev, \qquad  F_\pi=130.41(20)\mev,\qquad \frac{F_K}{F_\pi}=1.194(5)\, ,
\ee
\be
m_s(m_c)=109.1(2.8)\mev, \qquad  m_d(m_c)=5.44(19)\mev\,.
\ee

New sources of flavour and CP violation in 331 models are parametrized by
new mixing parameters and phases
\be\label{PAR}
\tilde s_{13},\qquad\tilde s_{23},\qquad  \delta_1,\qquad \delta_2
\ee
with $\tilde s_{13}$ and $\tilde s_{23}$ positive definite and 
$0\le \delta_{1,2}\le 2\pi$. They can be constrained by flavour observables as demonstrated in detail in \cite{Buras:2012dp}. 

 Noticeably, constraints deriving from $\Delta F=2$ observables do not depend on the choice of the fermion representation.  We recall here that, as already mentioned, the choice of the transformation properties of the fermions under the gauge group of 331 models, and in particular under the action of $SU(3)_L$, is not unique. Following \cite{Buras:2014yna} we denote by $F_1$ the fermion representation in which the first two generations of quarks transform as triplets under $SU(3)_L$ while the third one as well as leptons transform as antitriplets. On the other hand,  $F_2$ corresponds to  the case in which the choice of triplets and antitriplets is reversed. Right-handed fermions are always singlets.

\boldmath
\subsection{$Z^\prime$ Contribution}
\unboldmath
For the fermion representation $F_1$  we find
\be\label{C8prime}
C_8(m_c,Z^\prime)= 1.35 \left[\frac{g_2 s_W^2}{6c_W}\right]_{M_{Z^\prime}}\beta\sqrt{f(\beta)}\frac{\Delta_L^{sd}(Z^\prime)}{M_{Z^\prime}^2}= 4.09\times 10^{-2}
\beta\sqrt{f(\beta)}\frac{\Delta_L^{sd}(Z^\prime)}{M_{Z^\prime}^2}
\ee
with $1.35$ being renormalization group factor calculated for $M_{Z^\prime}=3\tev$ in \cite{Buras:2014yna}. $C_8(m_c,Z^\prime)$ depends through $\beta\sqrt{f(\beta)}$ on the 331 model considered where
\be\label{central}
f(\beta)=\frac{1}{1-(1+\beta^2)s_W^2} > 0\,.
\ee
We have indicated that the values of $g_2$ and $s_W^2$ should be evaluated 
at $M_{Z^\prime}$
with $s_W^2=\sin^2\theta_W=0.249$ and $g_2(M_{Z^\prime})=0.633$ corresponding to $M_{Z^\prime}=3\tev$.

The coupling $\Delta_L^{sd}(Z^\prime)$ is given in terms of the parameters 
in (\ref{PAR})
 as follows
\be\label{csd}
\Delta_L^{sd}(Z^\prime)=\frac{g_2(M_{Z^\prime})}{\sqrt{3}}c_W \sqrt{f(\beta)} \tilde s_{13}\tilde s_{23} e^{i(\delta_2-\delta_1)}\,.
\ee

The formulae  (\ref{C8prime})--(\ref{csd})  are valid for the fermion representation $F_1$. For a given 
$\beta$, the formulae for the fermion representation $F_2$ are obtained by 
reversing the sign in front of $\beta$. We note that $\Delta_L^{sd}(Z^\prime)$
is independent of the fermion representation as $f(\beta)$ depends only on $\beta^2$.

Combining all these formulae we find
\be\label{eprimeZP}
\left(\frac{\varepsilon'}{\varepsilon}\right)_{Z^\prime}= \pm 1.10 \,[\beta f(\beta)]\,
\tilde s_{13} \tilde s_{23} \sin(\delta_2-\delta_1)
\left[\frac{\bei}{0.76}\right]\left[\frac{3\tev}{M_{Z^\prime}}\right]^2
\ee
with the upper sign for $F_1$ and the lower for $F_2$ 

We observe that the contribution of $Z^\prime$ to $\epe$ is invariant under the transformation
\be\label{symmetry}
\beta \rightarrow -\beta, \qquad F_1 \rightarrow F_2\,.
\ee
This invariance is in fact valid for other flavour observables 
in the absence of $Z-Z^\prime$ mixing. But as pointed out in 
\cite{Buras:2014yna} in the presence of this  mixing it 
is broken as we will see soon.

\boldmath
\subsection{$Z$ Contribution}
\unboldmath
In the case of tree-level $Z$ contribution  also the operator $Q_8$ dominates but its Wilson coefficient
is given first by  \cite{Buras:2014sba}
\be\label{C80}
C_8(m_c,Z)= -0.76 \left[\frac{g_2 s_W^2}{6c_W}\right]_{M_Z}\frac{\Delta_L^{sd}(Z)}{M_{Z}^2}\,.
\ee
Here $g_2=g_2(M_Z)=0.652$ is the $SU(2)_L$ gauge coupling and the factor $0.76$ is the outcome of the RG evolution. 

In the 331 models the flavour violating couplings of $Z$ are generated through $Z-Z^\prime$ mixing. They are given by \cite{Buras:2014yna}
\be\label{ZZprime}
 \Delta^{ij}_L(Z) =\sin\xi \, \Delta^{ij}_L(Z^\prime) 
\qquad  (i\not=j) 
\ee
with 
\be\label{sxi}
\sin\xi=\frac{c_W^2}{3} \sqrt{f(\beta)}\left(3\beta \frac{s_W^2}{c_W^2}+\sqrt{3}a\right)\left[\frac{M_Z^2}{M_{Z^\prime}^2}\right]\, \equiv B(\beta,a) \left[\frac{M_Z^2}{M_{Z^\prime}^2}\right]
\ee
describing the $Z-Z^\prime$ mixing, where $s_W^2=0.23126$. 
 It should be stressed that this mixing is 
independent of the fermion representation. Here
\be\label{basica}
a=\frac{1-\tan^2\bar\beta}{1+\tan^2\bar\beta}, \qquad \tan\bar\beta=\frac{v_\rho}{v_\eta}.
\ee
As the vacuum expectation values of the Higgs triplets $\rho$ and 
$\eta$ are responsible for the masses of up-quarks and down-quarks, respectively,  we expressed the parameter $a$ in terms of the usual $\tan\bar\beta$ where 
we introduced a {\it bar} to distinguish the usual angle $\beta$  from the 
parameter $\beta$ in 331 models. See  \cite{Buras:2014yna} for further details.

We thus find 
\be\label{C8}
C_8(m_c,Z)= -0.76 \left[\frac{g_2 s_W^2}{6c_W}\right]_{M_Z} B(\beta,a)\frac{\Delta_L^{sd}(Z^\prime)}{M_{Z^\prime}^2}\,.
\ee

Combining these formulae allows to derive
a simple relation
\be\label{equ:Repsdef}
\left(\frac{\varepsilon'}{\varepsilon}\right)_{Z}=R_{\varepsilon^\prime}
\left(\frac{\varepsilon'}{\varepsilon}\right)_{Z^\prime},
\ee
where 
\be\label{equ:Repsprime}
R_{\varepsilon^\prime}=\frac{C_8(m_c,Z)}{C_8(m_c,Z^\prime)}= \mp \frac{0.53}{\beta}\, \frac{c_W^2}{3}\left[3\beta\frac{s_W^2}{c_W^2}+\sqrt{3}{a}\right] \,.
\ee
The upper sign in the expression (\ref{equ:Repsprime})
is valid for fermion representation $F_1$, the lower for $F_2$. 
The values of $R_{\varepsilon^\prime}$ are listed in Table~\ref{tab:Reps}
 for various values of $\beta$, $\tan\bar\beta$ and the representations $F_1$ 
and $F_2$.  Evidently $Z^\prime$ dominates 
NP contributions to $\epe$ implying that $Z-Z^\prime$ mixing effects are 
 small in this ratio. The two exceptions are the case of $\beta=-1/\sqrt{3}$ 
and $\tan\bar\beta=5$ and the  case of $\beta=1/\sqrt{3}$ 
and $\tan\bar\beta=0.2$ for which $Z$ contribution reaches $50\%$ of the $Z^\prime$ one. However, as seen in Table~\ref{tab:331models} both models are not 
among favourits and the largest $Z-Z^\prime$ effect of $25\%$ among the chosen models in that table is found in M6.

\begin{table}[!tb]
{\renewcommand{\arraystretch}{1.3}
\begin{center}
\begin{tabular}{|c||c|c|c|c|}
\hline
$\beta$ & $-\frac{2}{\sqrt{3}}$ & $-\frac{1}{\sqrt{3}}$ & $\frac{1}{\sqrt{3}}$ & $\frac{2}{\sqrt{3}}$ \\
\hline
$R_{\varepsilon^\prime}$\,($\tan\bar\beta=0.2$) & $\pm 0.066$ &$\pm 0.25$ &$\mp 0.50$ &$\mp 0.31$  \\
$R_{\varepsilon^\prime}$\,($\tan\bar\beta=1.0$) & $\mp 0.12$ &$\mp 0.12$ &$\mp 0.12$ &$\mp 0.12$  \\
$R_{\varepsilon^\prime}$ \,($\tan\bar\beta=5.0$) & $\mp 0.31$ &$\mp 0.50 $ &$\pm 0.25$ &$\pm 0.066$  \\
\hline
\end{tabular}
\end{center}}
\caption{\it $R_{\varepsilon^\prime}$ 
for different $\beta$ and $\tan\bar\beta$. The upper sign is for $F_1$, the lower for $F_2$.
\label{tab:Reps}}~\\[-2mm]\hrule
\end{table}

It should be noted that whereas the $Z^\prime$ contribution to $\epe$ for 
the representation $F_2$ differs from the one for $F_1$ by sign, the 
contribution of $Z$ to $\epe$ is independent of the fermion representation. 
This disparity breaks the  invariance in (\ref{symmetry}). Analogous 
feature is observed in several other flavour observables.

\boldmath
\subsection{Final Formula}
\unboldmath
The final expression for the shift of $\epe$ in 331 models is 
given by
\be\label{equ:Repstot}
\Delta (\epe) =(1+R_{\varepsilon^\prime})
\left(\frac{\varepsilon'}{\varepsilon}\right)_{Z^\prime}\,.
\ee
In the next section we will investigate, which 331 models can provide a 
significant shift $\Delta(\epe)$ for $M_{Z^\prime}=3\tev$ and how 
this shift is correlated with other flavour observables.

\section{Implications of Enhanced $\epe$ in 331 Models}\label{sec:3}
\subsection{Favourite 331 Models}
The favourite 331 models selected in \cite{Buras:2014yna} 
on the basis of their perfomance in electroweak precision tests  are listed 
in the notation of that paper
in Table~\ref{tab:331models}. In addition to the fermion representation and the values 
of $\beta$ and $\tan\bar\beta$ in a given model we indicate how in that 
model NP effects in the branching ratio for $B_s\to\mu^+\mu^-$ are correlated 
with the ones in $C_9$. For  $\mathcal{B}(B_s\to\mu^+\mu^-)$ the signs $\pm$ denote the enhancement and suppression of it with respect to its SM value, respectively.
$C_9$ in the SM is positive and $\pm$ also here denote the enhancements and 
suppressions  with respect to its SM value, respectively. In M6 and M11 NP 
contributions to $C_9$ are fully negligible. These correlations are shown in 
Fig.~15 of  \cite{Buras:2014yna}.

\begin{table}[!tb]
{\renewcommand{\arraystretch}{1.3}
\begin{center}
\begin{tabular}{|c||c|c|c||c|c|c|}
\hline
MI  &     {\rm scen.} &  $\beta$ & $\tan {\bar \beta}$ & $\mathcal{B}(B_s\to\mu^+\mu^-)$ & $C_9$ &$\sin(\delta_2-\delta_1)$\\
\hline
M3 & $F_1$ & $-1/\sqrt{3}$ & 1 & $\pm$ & $\pm$& $-1$ \\
M6 & $F_1$ & $1/\sqrt{3}$ & 5 & $\pm$  & $0$ &$+1$ \\
M8 & $F_1$ & $2/\sqrt{3}$ & 5 &$\pm$ & $\mp$& $+1$ \\
M9 & $F_2$ & $-2/\sqrt{3}$ & 1 & $\pm$ & $\mp$&$+1$ \\
M11 & $F_2$ & $-1/\sqrt{3}$ & 1 & $\pm$  & $0$&$+1$ \\
M14 & $F_2$ & $1/\sqrt{3}$ & 5 &$\pm$ & $\pm$&$-1$ \\
M16 & $F_2$ & $2/\sqrt{3}$ & 5 & $\pm$ & $\pm$&$-1$
\\
\hline
\end{tabular}
\end{center}}
\caption{\it Definition of the favourite 331 models. See text for explanation of the columns for $B_s\to\mu^+\mu^-$ and $C_9$. In the last column we list the
values of $\sin(\delta_2-\delta_1)$ for which the maximal {\it positive} shifts 
of $\epe$ in a given model can be obtained.
\label{tab:331models}}~\\[-2mm]\hrule
\end{table}

Before entering the discussion of $\epe$ let us recall that the present 
data favour simultaneous suppressions of  $\mathcal{B}(B_s\to\mu^+\mu^-)$ 
and $C_9$. From Fig.~15 in \cite{Buras:2014yna} and Table~\ref{tab:331models}
we reach the following conclusions.

\begin{itemize}
\item
Qualitatively models M3, M14 and M16 can provide simultaneous suppression
of $\mathcal{B}(B_s\to\mu^+\mu^-)$ and a
negative shift  ${\rm Re} C_9^\text{NP}$ but the suppression of 
 $\mathcal{B}(B_s\to\mu^+\mu^-)$  is not significant.
\item
For softening the $B_d\to K^*\mu^+\mu^-$ anomaly the most interesting is the 
model M16. 
If the anomaly in question  remains but decreases with time also models M3 and
M14 would be of interest.
\item
The remaining four models, in fact the four top models on our list of 
favourites in  (\ref{favoured}) below, as far as electroweak precision tests are 
concerned, do not provide any explanation of  
$B_d\to K^*\mu^+\mu^-$ anomaly but are interesting for $B_{s}\to\mu^+\mu^-$ 
decay. These are {\rm M6},  {\rm M8}, {\rm M9} and {\rm M11}, the first two 
with $F_1$ and the last two with $F_2$ fermion representation. It turns out that the strongest suppression of the rate for 
 $B_{s}\to\mu^+\mu^-$ can be achieved in M8 and M9. In fact these two models are the two leaders on the list of 
favourites in  (\ref{favoured}).  But in these models $C_9$ is enhanced 
and not suppressed as presently observed in the data. The suppression of the $B_s\to\mu^+\mu^-$ rate is smaller in M6 and M11 but there the shift in $C_9$ 
can be neglected.
\end{itemize}

We conclude that when the data for $\mathcal{B}(B_s\to\mu^+\mu^-)$ and $C_9$  improve we will be able to reduce the number of favourite models. But if both will be 
significantly suppressed none of the models considered here will be able to 
describe the data. In fact model M2 with $F_1$, $\beta=-2/\sqrt{3}$ and $\tan\bar\beta=5$ could in principle do this work here but it is disfavoured through 
electroweak precision tests.

Concerning these tests the ranking is given as follows
\be\label{favoured}
{\rm M9}, \quad {\rm M8},\quad {\rm M6}, \quad {\rm M11}, \quad {\rm M3}, \quad {\rm M16}, \quad {\rm M14}, \qquad {(\rm favoured)}
\ee
with the first five performing better than the SM while the last two 
 basically as the SM.
The models with {\it odd} index I correspond to $\tan\bar\beta=1.0$ and 
the ones with {\it even} one to  $\tan\bar\beta=5.0$.   None of the 
models with  $\tan\bar\beta=0.2$ made this list implying reduced impact of
$Z-Z^\prime$ mixing on $\epe$ and small NP effects in decays with 
neutrinos in the final state.

\boldmath
\subsection{Predictions for $\epe$ in Favourite Models}
\unboldmath
After the recollection of the correlations among $B$ physics observables in 
the seven models in questions we are in the position to investigate which 
of these models allows for significant enhancement of $\epe$.

To this end we set the CKM parameters to
\be\label{CKMfix}
\vub=3.6\times 10^{-3}, \qquad \vcb=42.0 \times 10^{-3}, \qquad \gamma=70^\circ.
\ee
This choice is in the ballpark  of present best values for these three 
parameters but is also motivated by the fact that NP contributions to 
$\varepsilon_K$ in 331 models are rather small for $M_{Z^\prime}$ of a few $\tev$ and SM 
should perform well in this case. Indeed for this choice of CKM parameters 
we find
\be
|\varepsilon_K|_{\rm SM}=2.14\times 10^{-3}, \qquad (\Delta M_K)_{\rm SM}=0.467\cdot 10^{-2} \,\text{ps}^{-1}
\ee
and  $|\varepsilon_K|$ in 
the SM only $4\%$ below the data. Due to the presence of long distance effects
in $\Delta M_K$ also this value is compatible with the data.
Moreover, the resulting ${\rm Im}\lambda_t=1.42 \times 10^{-4}$ is very close to the central value  ${\rm Im}\lambda_t=1.40 \times 10^{-4}$ used in \cite{Buras:2015yba}. While our choice of CKM parameters is irrelevant 
for the shift in $\epe$ it matters in the predictions for NP contributions to rare $K$ and $B$ decays due to the their intereference with  SM  contributions.

Next, as in \cite{Buras:2012dp}, we  perform a simplified analysis of
$\Delta M_{d,s}$, $S_{\psi K_S}$
and $S_{\psi\phi}$
in order to identify oases in the space of four parameters in (\ref{PAR})
for which these four observables are consistent with experiment.
To this end we use the formulae for $\Delta F=2$ observables in 
 \cite{Buras:2012dp,Buras:2014yna} and 
 set all input parameters listed in Table~\ref{tab:input} at their central values. But in order to take partially hadronic
and experimental uncertainties into account we require the 331 models
to reproduce the data for $\Delta M_{s,d}$ within $\pm 10\%(\pm 5\%)$ and the
data on $S_{\psi K_S}$ and $S_{\psi\phi}$ within experimental
$2\sigma$. As seen in  Table~\ref{tab:input} the present uncertainties in 
hadronic parameters relevant for $\Delta M_{s,d}$ are larger than $10\%$ but
we anticipate progress in the coming years. The accuracy of $\pm 5\%$ should 
be achieved at the end of this decade.

Specifically, our search is governed by the following allowed ranges:
\be
16.0\, (16.9)/{\rm ps}\le \Delta M_s\le 19.5\, (18.7)/{\rm ps},
\quad  -0.055\le S_{\psi\phi}\le 0.085, \label{oases23}
\ee
\be
0.46\, (0.48)/{\rm ps}\le \Delta M_d\le 0.56\, (0.53)/{\rm ps},\quad  0.657\le S_{\psi K_S}\le 0.725\, , \label{oases13}
\ee
where the values in parentheses correspond to decreased uncertainty. For the
central parameters we find in the SM
\be
 (\Delta M_s)_{\rm SM}=18.45/{\rm ps},\qquad (\Delta M_d)_{\rm SM}=0.558/{\rm ps},
\qquad S^{\rm SM}_{\psi\phi}=0.037, \qquad S^{\rm SM}_{\psi K_S}=0.688\,.
\ee

In the case of $\varepsilon_K$ the status of hadronic parameters is better than 
for $\Delta M_{s,d}$ but the CKM uncertainties are larger  and the result depends  on whether inclusive or exclusive determinations of $\vub$ and 
$\vcb$ are used. As we keep these parameters fixed we include this uncertainty 
by choosing the allowed range for  $|\varepsilon_K|$ below to be roughly the 
range one would get in the SM by varying $\vub$ and $\vcb$ in their ranges 
known from tree-level determinations.

The uncertainties in 
$\Delta M_K$ are very large both due to the presence of long distance effects 
and large uncertainty in $\eta_{cc}$. We could in principle ignore this constraint but as we will see in the next section it plays a role for  $M_{Z^\prime}$ 
above $30\tev$ not allowing for large shifts in $\epe$ in 331 models for 
such high values of  $M_{Z^\prime}$. In fact as we will explain in the next 
section it is $\Delta M_K$ and not $\varepsilon_K$ which is most constraining 
the maximal values of $\epe$. But at the LHC scales and even at 
 $M_{Z^\prime}=10\tev$ the $\Delta M_K$ constraint is irrelevant. Only 
for scales above $M_{Z^\prime}=30\tev$ it starts to play an important role bounding the maximal values of $\epe$.  Once the knowledge of long distance effects improves and the error on  $\eta_{cc}$ decreases it will be possible to improve our analysis in this part.

We will then 
impose the ranges
\be\label{CONE}
1.60\times 10^{-3}< |\varepsilon_K|< 2.50\times 10^{-3}\,, \qquad
-0.30\le \frac{(\Delta M_K)^{Z^\prime}}{(\Delta M_K)_\text{exp}}\le 0.30\,.
\ee

\begin{table}[!tb]
\center{\begin{tabular}{|l|l|}
\hline
$G_F = 1.16638(1)\cdot 10^{-5}\gev^{-2}$\hfill\cite{Agashe:2014kda} 	&  $m_{B_d}=5279.58(17)\mev$\hfill\cite{Agashe:2014kda}\\
$M_W = 80.385(15) \gev$\hfill\cite{Agashe:2014kda}  								&	$m_{B_s} =
5366.8(2)\mev$\hfill\cite{Agashe:2014kda}\\
$\sin^2\theta_W = 0.23126(13)$\hfill\cite{Agashe:2014kda} 				& 	$F_{B_d} =
190.5(42)\mev$\hfill \cite{Aoki:2013ldr}\\
$\alpha(M_Z) = 1/127.9$\hfill\cite{Agashe:2014kda}									& 	$F_{B_s} =
227.7(45)\mev$\hfill \cite{Aoki:2013ldr}\\
$\alpha_s(M_Z)= 0.1185(6) $\hfill\cite{Agashe:2014kda}			&  $\hat B_{B_d} =1.27(10)$,  $\hat
B_{B_s} =
1.33(6)$\;\hfill\cite{Aoki:2013ldr}
\\\cline{1-1}
$m_d(2\gev)=4.68(16)\mev$	\hfill\cite{Aoki:2013ldr}						& 
$\hat B_{B_s}/\hat B_{B_d}
= 1.06(11)$ \hfill \hfill\cite{Aoki:2013ldr} \\
$m_s(2\gev)=93.8(24) \mev$	\hfill\cite{Aoki:2013ldr}							& 
$F_{B_d} \sqrt{\hat
B_{B_d}} = 216(15)\mev$\hfill\cite{Aoki:2013ldr} \\
$m_c(m_c) = 1.275(25) \gev$ \hfill\cite{Agashe:2014kda}	
				&
$F_{B_s} \sqrt{\hat B_{B_s}} =
266(18)\mev$\hfill\cite{Aoki:2013ldr} \\
$m_b(m_b)=4.18(3)\gev$\hfill\cite{Agashe:2014kda} 				&
$\xi =
1.268(63)$\hfill\cite{Aoki:2013ldr} \\
  $m_t(m_t) = 163(1)\gev$\hfill\cite{Allison:2008xk}     			& $\eta_B=0.55(1)$\hfill\cite{Buras:1990fn,Urban:1997gw}
\\\cline{1-1}
$m_K= 497.614(24)\mev$	\hfill\cite{Agashe:2014kda}
						&  $\Delta M_d = 0.510(3)
\,\text{ps}^{-1}$\hfill\cite{Amhis:2014hma}\\ 
$F_K/F_\pi=1.194(5)$\hfill\cite{Agashe:2014kda}								&  $\Delta M_s = 17.757(21)
\,\text{ps}^{-1}$\hfill\cite{Amhis:2014hma}
\\	
$F_\pi = 130.41(20)\mev$\hfill\cite{Agashe:2014kda}													&
$S_{\psi K_S}= 0.691(17)$\hfill\cite{Amhis:2014hma}\\
$\hat B_K= 0.750(15)$\hfill \cite{Aoki:2013ldr,Buras:2014maa}										&
$S_{\psi\phi}= 0.015(35)$\hfill \cite{Amhis:2014hma}\\
$\kappa_\epsilon=0.94(2)$\hfill\cite{Buras:2008nn,Buras:2010pza}				& $\Delta\Gamma_s/\Gamma_s=0.122(9)$\hfill\cite{Amhis:2014hma}
\\	
$\eta_{cc}=1.87(76)$\hfill\cite{Brod:2011ty}												
	& $\tau_{B_s}= 1.509(4)\,\text{ps}$\hfill\cite{Amhis:2014hma}\\		
$\eta_{tt}=0.5765(65)$\hfill\cite{Buras:1990fn}												
& $\tau_{B_d}= 1.520(4) \,\text{ps}$\hfill\cite{Amhis:2014hma}\\
$\eta_{ct}= 0.496(47)$\hfill\cite{Brod:2010mj}												
& $\tau_{B^\pm}= 1.638(4)\,\text{ps}$\hfill\cite{Amhis:2014hma}    \\
$\Delta M_K= 0.5293(9)\cdot 10^{-2} \,\text{ps}^{-1}$\hfill\cite{Agashe:2014kda}	&
$|V_{us}|=0.2253(8)$\hfill\cite{Agashe:2014kda}\\
$|\eps_K|= 2.228(11)\cdot 10^{-3}$\hfill\cite{Agashe:2014kda}					& $\gamma =(73.2^{+6.3}_{-7.0})^\circ$
\hfill\cite{Trabelsi:2014}\\
\hline
\end{tabular}  }
\caption {\textit{Values of the experimental and theoretical
    quantities used as input parameters  as of June 2015. For future 
updates see PDG \cite{Agashe:2014kda}, FLAG \cite{Aoki:2013ldr} and HFAG \cite{Amhis:2014hma}. }}
\label{tab:input}
\end{table}

The search for the oases in question is simplified by the fact that the pair
$(\Delta M_s,S_{\psi\phi})$ depends only on $(\tilde s_{23},\delta_2)$,
while the pair
$(\Delta M_d,S_{\psi K_S})$ only on $(\tilde s_{13},\delta_1)$. The result of 
this search is similar to the one found in Figs.~5 and 6 in \cite{Buras:2012dp}
but the oases differ in details because of slight changes in input parameters 
and the reduced allowed range on $S_{\psi\phi}$ by about  a factor of three. 

\begin{figure}[!tb]
 \centering
\includegraphics[width = 0.45\textwidth]{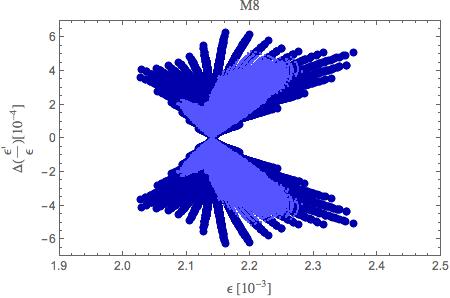}
\includegraphics[width = 0.45\textwidth]{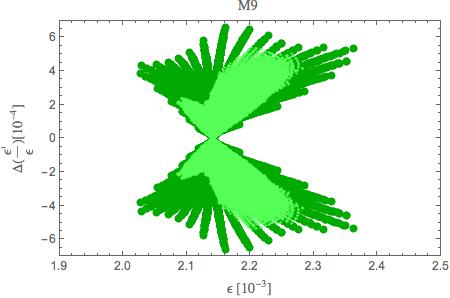}
\includegraphics[width = 0.45\textwidth]{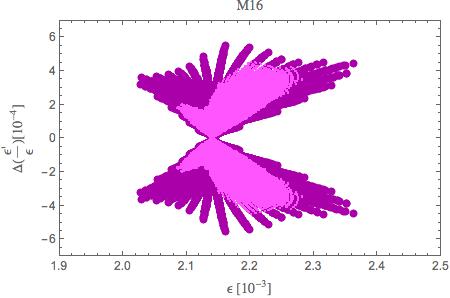}
\caption{ \it $\Delta(\epe)$ versus $\varepsilon_K$ for M8, M9 and M16. $M_{Z^\prime}=3\tev$ . Darker regions correspond to  present constraints on $\Delta M_{s,d}$ and lighter ones to future projection.
}\label{M8M9M16}~\\[-2mm]\hrule
\end{figure}

\begin{figure}[!tb]
 \centering
\includegraphics[width = 0.45\textwidth]{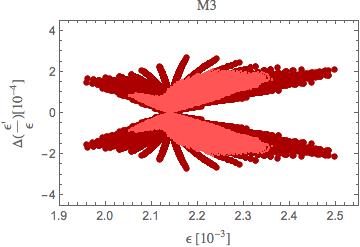}
\includegraphics[width = 0.45\textwidth]{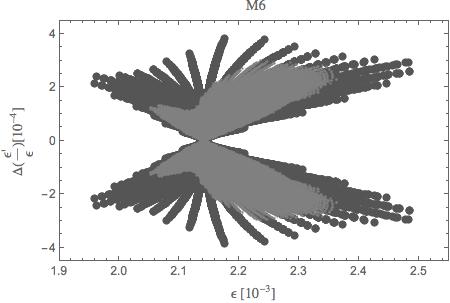}
\includegraphics[width = 0.45\textwidth]{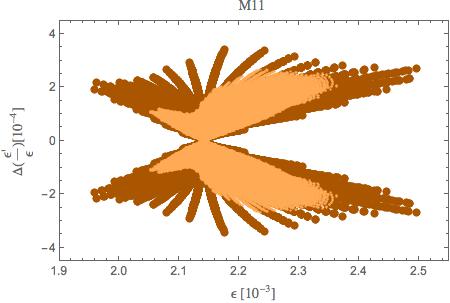}
\includegraphics[width = 0.45\textwidth]{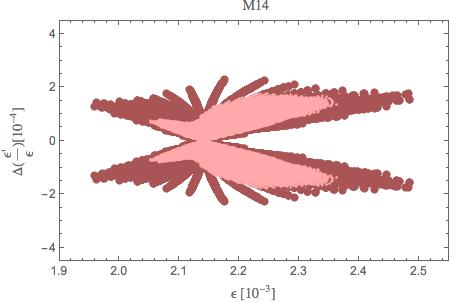}
\caption{$\Delta(\epe)$ versus $\varepsilon_K$ for  M3, M6, M11 and M14. $M_{Z^\prime}=3\tev$ . Darker regions correspond to  present constraints on $\Delta M_{s,d}$ and lighter ones to future projection. 
}\label{REST}~\\[-2mm]\hrule
\end{figure}

Having determined the ranges for the parameters (\ref{PAR}) we can calculate 
all the remaining flavour observables of interest. In particular we can eliminate those models  listed in Table~\ref{tab:331models} which are not capable 
of providing a shift in $\epe$ larger than say $4\times 10^{-4}$. To this 
end we show in Fig.~\ref{M8M9M16} this shift as a function of $\varepsilon_K$ for models 
M8, M9 and M16 and in  Fig.~\ref{REST} this shift for the remaining models. On the basis
of these results we observe the following:
\begin{itemize}
\item
Only models M8, M9 and M16 are of interest to us as far as $\epe$ is concerned 
and in what follows we will concentrate our numerical analysis on these three
models.
\item
Interestingly, as mentioned above, the strongest suppression of the rate for 
 $B_{s}\to\mu^+\mu^-$ can be achieved in M8 and M9 although they have presently 
difficulties with the LHCb anomalies. Using the formulae in  \cite{Buras:2014yna} this can be expressed in terms of 
the relations between the coefficients $C_9^{\rm NP}$ and  $C_{10}^{\rm NP}$ 
which are independent of $M_{Z^\prime}$ and read 
\be
 C_9^{\rm NP}=0.51\, C_{10}^{\rm NP}\quad ({\rm M8}) \qquad 
 C_9^{\rm NP}=0.42\, C_{10}^{\rm NP}\quad ({\rm M9})\,.
\ee
They differ significantly from the favourite relations $C_9^{\rm NP}=- C_{10}^{\rm NP}$ or $C_9^{\rm NP}\gg C_{10}^{\rm NP}$ \cite{Altmannshofer:2014rta,Descotes-Genon:2015uva}.
\item
On the other hand M16 is the most interesting model for softening the $B_d\to K^*\mu^+\mu^-$ anomaly but cannot help by much in suppressing $B_s\to \mu^+\mu^-$.
One finds in this case
\be
C_9^{\rm NP}=-4.61\, C_{10}^{\rm NP}\quad ({\rm M16}) 
\ee
which is much closer to one of the favourite solutions in which NP resides dominantly in the 
coefficient $C_9$.
\item
Thus already on the basis of $B$ physics observables we should be able to 
distinguish between (M8,M9) and M16. But the common feature of the three models 
is that they provide a bigger shift in $\epe$ when the SM value of $\varepsilon_K$ is below the data and a positive shift in  $\varepsilon_K$  is required. This
is in particular seen in the case of lighter colours describing decreased uncertainties in $\Delta M_{s,d}$.
\item
Most importantly positive shifts in $\epe$ in the ballpark of $6\times 10^{-4}$ are possible in these three models, but they are somewhat reduced when the
allowed range for $\Delta M_{s,d}$ is reduced. Such shifts could be in principle sufficient to describe the data for $\epe$.
\end{itemize}

The question then arises whether the models M8 and M9 while enhancing $\epe$ 
can simultaneously suppress the rate for $B_{s}\to\mu^+\mu^-$ and whether M16 
while enhancing $\epe$ can simultaneously suppress ${\rm Re} C_9$. Moreover 
correlation of $\Delta(\epe)$ with the branching ratios for  $\kpn$ and $\klpn$
is of great interest in view of NA62 and KOPIO experiments.

In order to answer these questions we  used the formulae for 
$B_{s,d}\to\mu^+\mu^-$, $C_9$, $\kpn$ and $\klpn$ 
presented in  \cite{Buras:2014yna} together with the formulae for $\epe$ 
collected in the previous section to calculate for M8, M9 and M16 models
correlation of $\Delta (\epe)$ with 
$\mathcal{B}(B_s\to\mu^+\mu^-)$, ${\rm Re} C^{\rm NP}_9$, $\mathcal{B}(\kpn)$
and $\mathcal{B}(\klpn)$. The result is shown in Figs.~\ref{CORR8}, \ref{CORR9} 
and \ref{CORR16}, respectively.

\begin{figure}[!tb]
 \centering
\includegraphics[width = 0.45\textwidth]{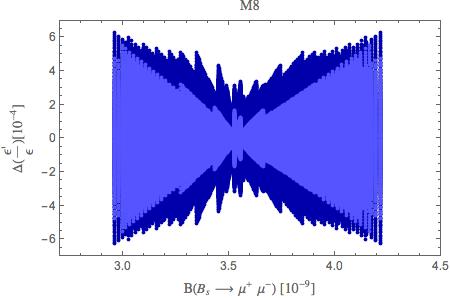}
\includegraphics[width = 0.45\textwidth]{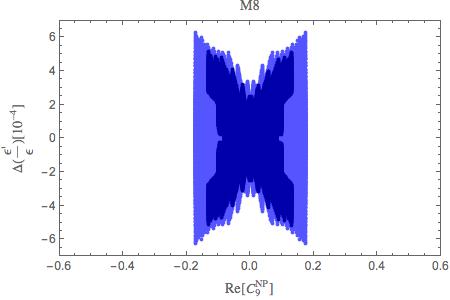}
\includegraphics[width = 0.45\textwidth]{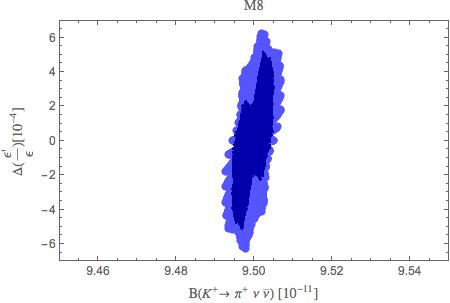}
\includegraphics[width = 0.45\textwidth]{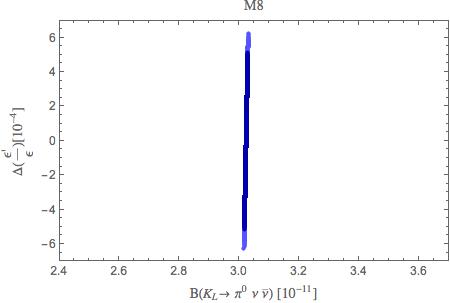}
\caption{ \it Correlations of $\Delta(\epe)$ with various observables  for M8 
at  $M_{Z^\prime}=3\tev$ . Colour coding  as in Fig~\ref{M8M9M16}.
}\label{CORR8}~\\[-2mm]\hrule
\end{figure}

\begin{figure}[!tb]
 \centering
\includegraphics[width = 0.45\textwidth]{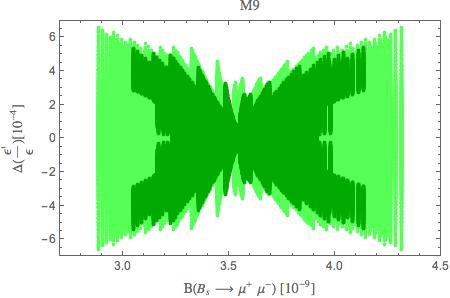}
\includegraphics[width = 0.45\textwidth]{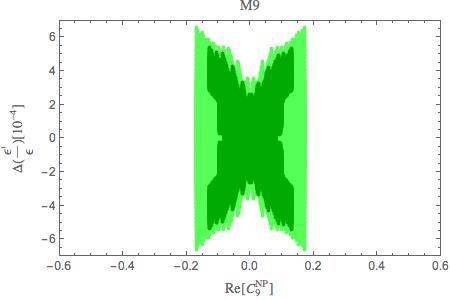}
\includegraphics[width = 0.45\textwidth]{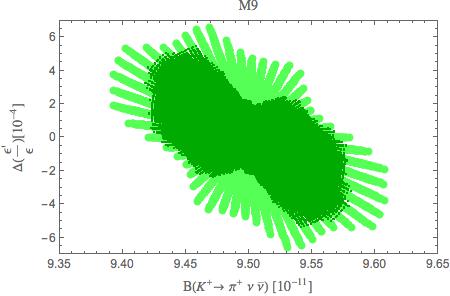}
\includegraphics[width = 0.45\textwidth]{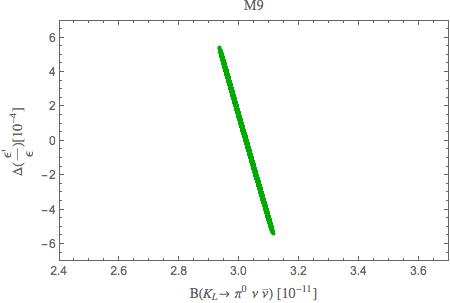}
\caption{ \it Correlations of $\Delta(\epe)$ with various observables  for M9 
at  $M_{Z^\prime}=3\tev$ .  Colour coding  as in Fig~\ref{M8M9M16}.
}\label{CORR9}~\\[-2mm]\hrule
\end{figure}

\begin{figure}[!tb]
 \centering
\includegraphics[width = 0.45\textwidth]{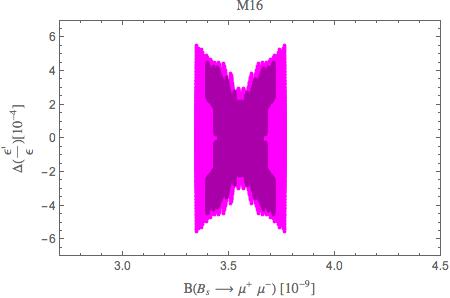}
\includegraphics[width = 0.45\textwidth]{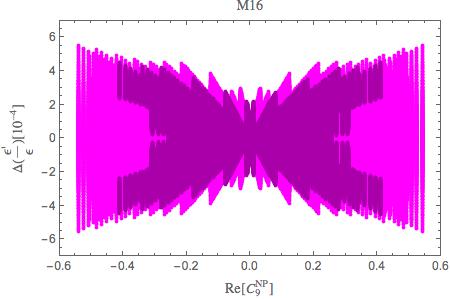}
\includegraphics[width = 0.45\textwidth]{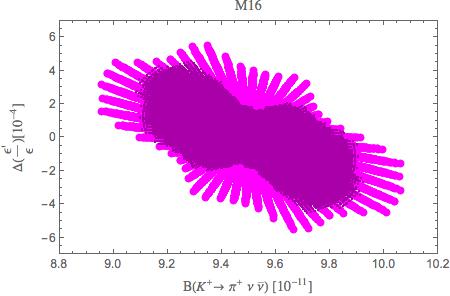}
\includegraphics[width = 0.45\textwidth]{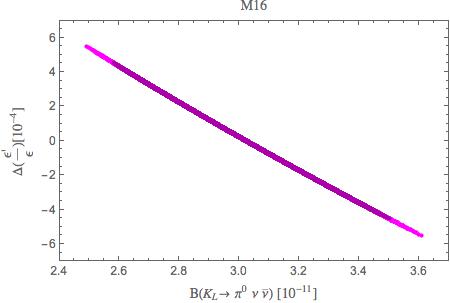}
\caption{ \it Correlations of $\Delta(\epe)$ with various observables  for M16
at  $M_{Z^\prime}=3\tev$ .  Colour coding  as in Fig~\ref{M8M9M16}.
}\label{CORR16}~\\[-2mm]\hrule
\end{figure}

We observe:
\begin{itemize}
\item 
Models M8 and M9 have similar pattern of deviations from the SM with NP effects only relevant in 
$\epe$ and $B_s\to\mu^+\mu^-$ and in fact positive shift of  $\epe$ up to 
$6\times 10^{-4}$ and simultaneous suppression 
of the rate for  $B_s\to\mu^+\mu^-$ up to $(15-20)\%$ are  possible in both 
models. NP effects in $C_9$, $\kpn$ and $\klpn$ are very small. 
\item
In M16 NP effects are only relevant in $\epe$, $C_9$ and $\klpn$. A 
 positive shift of  $\epe$ up to $5\times 10^{-4}$ and simultaneous negative shift of $C_9$ up to $-0.55$ is possible in this model. But then the rate for 
$\klpn$ is predicted to be suppressed by $15\%$ and the one for $\kpn$ by roughly $5\%$.
\end{itemize}

We conclude therefore that the distinction between M8 and M9 will be very difficult on the basis of observables considered by us but the distinction between these models and M16 should be possible  with  improved measurements of 
$B_s\to\mu^+\mu^-$ and improved determination of NP contribution to $C_9$ which is expected in the flavour precision era. But if NA62 collaboration finds 
the rate for $\kpn$ significantly above its SM value all these models will fail
in describing the data. The same applies to the models in Fig.~\ref{REST}.

\boldmath
\section{$Z^\prime$ Outside the Reach of the LHC}\label{sec:4}
\unboldmath
We will next investigate how the pattern of NP effects changes if 
 $M_{Z^\prime}$ is above $5\tev$ and out of the reach of the LHC. As already 
pointed out in  \cite{Buras:2012dp}, with increased $M_{Z^\prime}$ NP effects in 
$\varepsilon_K$ and $\Delta M_K$ increase relative to the ones in $\Delta M_{s,d}$ in 331 models due to particular structure of flavour violating couplings in
these models. While the coupling $\Delta_L^{sd}(Z^\prime)$ in (\ref{csd}) is proportional to the  product $\tilde s_{13}\tilde s_{23}$, the corresponding couplings relevant for $B_{s,d}$ system are given by
\begin{align}
\Delta_L^{bd}(Z^\prime)& =\frac{g_2(M_{Z^\prime})}{\sqrt{3}}c_W \sqrt{f(\beta)} \tilde s_{13} e^{-i\delta_1}\,,\\
\Delta_L^{bs}(Z^\prime)& =\frac{g_2(M_{Z^\prime})}{\sqrt{3}}c_W \sqrt{f(\beta)} \tilde s_{23} e^{-i\delta_2}\,,
\end{align}
each involving only one small parameter $\tilde s_{ij}$. $\Delta M_d$ and the 
CP asymmetry $S_{\psi K_S}$ specify the allowed ranges for $s_{13}$ and 
$\delta_1$, while $\Delta M_s$ and the 
CP asymmetry $S_{\psi \phi}$ specify the allowed ranges for $s_{23}$ and 
$\delta_2$. In this manner for a given $M_{Z^\prime}$ the allowed ranges of the 
four parameters entering the $K$ meson system are determined. But, can be 
further constrained by $\varepsilon_K$  and in particular by $\Delta M_K$ for sufficiently large  $M_{Z^\prime}$ as explained below. We refer to the plots in \cite{Buras:2012dp}.

In order to proceed we would like to point out that with increasing 
 $M_{Z^\prime}$  the RG analysis leading to (\ref{eprimeZP}) has to be improved 
modifying this formula to 
\be\label{eprimeZPlarge}
\left(\frac{\varepsilon'}{\varepsilon}\right)_{Z^\prime}= \pm r_{\varepsilon^\prime} 1.1 \,[\beta f(\beta)]\,
\tilde s_{13} \tilde s_{23} \sin(\delta_2-\delta_1)
\left[\frac{\bei}{0.76}\right]\left[\frac{3\tev}{M_{Z^\prime}}\right]^2
\ee
with the upper sign for $F_1$ and the lower for $F_2$. The parameter $r_{\varepsilon^\prime}$ takes into account additional RG evolution above $\mu=M_{Z^\prime}=3\tev$ into account and reaches $r_{\varepsilon^\prime}= 1.45$ for   $M_{Z^\prime}=100\tev$. This could turn out to be useful in models in which the $\Delta F=2$ 
constraints could be eliminated, for instance in the presence of other operators. But in 331 models this is not possible and as we will see for  $M_{Z^\prime}\ge 50\tev$ NP effects in $\epe$ are suppressed in all 331 models.  In Table~\ref{RG} we give the values of  $r_{\varepsilon^\prime}$ for different $M_{Z^\prime}$.

\begin{table}[!htb]
\begin{center}
\begin{tabular}{|c|c|c|c|c|c|c|}
\hline
  $M_{Z^\prime}$ & $3\tev$ & $6\tev$& $10\tev$& $20\tev$ & $50\tev$ & $100\tev$\\
\hline
 $r_{\varepsilon^\prime}$  & $1.00$ & $1.08$ & $1.15$ & $1.24$ &  $1.36$ & $1.45$ \\
\hline
\end{tabular}
\end{center}
\caption{The $M_{Z^\prime}$ dependence of $r_{\varepsilon^\prime}$.
\label{RG}}
\end{table}

In order to analyze NP effects beyond the LHC scales we recall the formulae 
for the shifts due to NP in $\Delta F=2$ observables.  

In the $K$ meson system we have
\be\label{eK}
(\Delta \varepsilon_K)^{Z^\prime}=1.76\times 10^4\, r_{\varepsilon}\left[\frac{3\tev}{M_{Z^\prime}}\right]^2 
 {\rm Im}\left[\Delta_L^{s d}(Z^\prime)^*\right]^2 \,.
\ee

\be\label{Mk}
\frac{(\Delta M_K)^{Z^\prime}}{(\Delta M_K)_\text{exp}}=
5.29\times 10^4\, r_{\varepsilon} \left[\frac{3\tev}{M_{Z^\prime}}\right]^2 
 {\rm Re}\left[\Delta_L^{s d}(Z^\prime)^*\right]^2 \,,
\ee
where $r_{\varepsilon}$ describes RG effects above $M_{Z^\prime}=3\tev$. These effects are much smaller than in the case of $\epe$ and in fact suppress slightly 
NP contribution to $\varepsilon_K$, $\Delta M_K$ and also $\Delta M_{s,d}$. 
But even for  $M_{Z^\prime}=100\tev$ this factor amounts to $r_{\varepsilon}\approx 0.95$ in the NP contributions to these observables and for  $M_{Z^\prime}\le 50\tev$ this effect can be fully neglected. But we keep this factor in formulae below for the
future in case various uncertainties decrease.

From (\ref{eprimeZPlarge}) and (\ref{eK}) we find
\be\label{triple}
(\Delta \varepsilon_K)^{Z^\prime} =
\mp \frac{g^2(M_{Z^\prime}) c_W^2}{\beta}
\tilde s_{13} \tilde s_{23} \cos(\delta_2-\delta_1)\left[\frac{r_{\varepsilon}}{r_{\varepsilon^\prime}}\right]
\left[\frac{0.76}{\bei}\right] 1.07 \times 10^4\, \left(\frac{\varepsilon'}{\varepsilon}\right)_{Z^\prime}
\ee
with $-$ for $F_1$ and $+$ for $F_2$. From (\ref{eprimeZPlarge}) and (\ref{Mk}) 
we have on the other hand
\be\label{triple2}
\frac{(\Delta M_K)^{Z^\prime}}{(\Delta M_K)_\text{exp}} =
\pm \frac{g^2(M_{Z^\prime}) c_W^2}{\beta}
\tilde s_{13} \tilde s_{23} \left[\frac{\cos(2(\delta_2-\delta_1))}{\sin(\delta_2-\delta_1)}\right] \left[\frac{r_{\varepsilon}}{r_{\varepsilon^\prime}}\right]
\left[\frac{0.76}{\bei}\right] 1.60\times 10^4\, \left(\frac{\varepsilon'}{\varepsilon}\right)_{Z^\prime}
\ee
with $+$ for $F_1$ and $-$ for $F_2$.
The SM contribution  to $\Delta M_K$ is subject to much larger hadronic uncertainties than is the case of $\varepsilon_K$ and it is harder to find out what is
the room left for NP contributions. In fact we do not even know the 
required sign of this contribution. But as we will see soon this constraint 
could become relevant with improved theory for high values of $M_{Z^\prime}$ and 
we will impose the constraint in (\ref{CONE}).
 For the present discussion we neglect the effects of 
$Z-Z^\prime$ mixing which will be included in the numerics.

The formula (\ref{triple}) represents a correlation between $\Delta M_{s,d}$,  $S_{\psi K_S}$ 
and  $S_{\psi \phi}$ which for a given $\beta$ determine all parameters on its r.h.s and the ratio of $\epe$ and 
$(\Delta \varepsilon_K)^{Z^\prime}$.
Even if $M_{Z^\prime}$ does not enter explicitly this expression, the allowed 
values for $\tilde s_{13}$ and $\tilde s_{23}$ depend on it. Similar 
comments apply to (\ref{triple2}).

The relation between the $Z'$ effects in $\Delta F=2$
master functions $S_i$ are related as follows \cite{Buras:2012dp}
\be\label{master1}
\frac{\Delta S_K}{\Delta S_d\Delta S_s^*}\propto M^2_{Z'} \left[\frac{\Delta_L^{sd}(Z')}{\Delta_L^{bd}(Z')\Delta_L^{bs*}(Z')}\right]^2 \propto
 M^2_{Z'}\,(1-(1+\beta^2)s_W^2).
\ee
Indeed NP contributions to $\Delta M_{s,d}$ are proportional to $\Delta S_{s,d}$ and in $\varepsilon_K$ to 
$\Delta S_K$. This relation follows then from the fact that 
\be
\Delta S_s \propto \left[\frac{\tilde s_{23}}{M_{Z^\prime}}\right]^2, \qquad 
\Delta S_d \propto \left[\frac{\tilde s_{13}}{M_{Z^\prime}}\right]^2.\qquad
\Delta S_K \propto \left[\frac{\tilde s_{13}\tilde s_{23}}{M_{Z^\prime}}\right]^2.
\ee

Presently the strongest constraints on the parameters $\tilde s_{ij}$ come 
from $\Delta F=2$ processes. With increasing value of $M_{Z^\prime}$ the maximal 
values of  $\tilde s_{ij}$ allowed by these constraints increase with increasing $M_{Z^\prime}$. This in turn has impact on the   $M_{Z^\prime}$ dependence of maximal values of NP contributions to $\Delta F=1$ observables that in addition to
explicit   $M_{Z^\prime}$ dependence through $Z^\prime$ propagator depend sensitively on $\tilde s_{ij}$. 

\begin{figure}[!tb]
 \centering
\includegraphics[width = 0.45\textwidth]{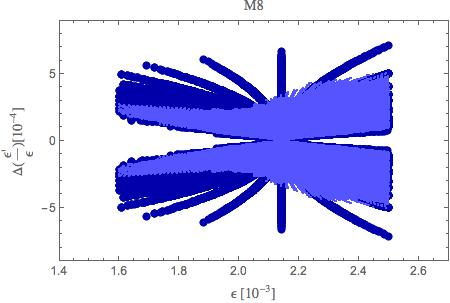}
\includegraphics[width = 0.45\textwidth]{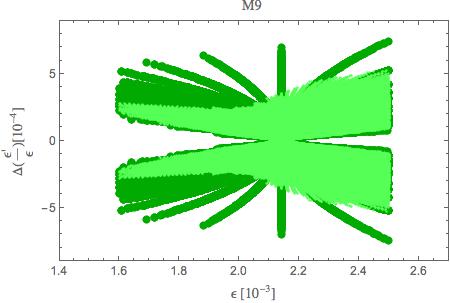}
\includegraphics[width = 0.45\textwidth]{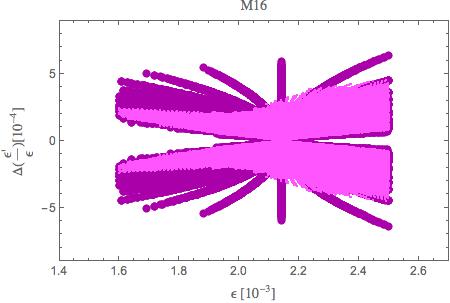}
\caption{ \it $\Delta(\epe)$ versus $\varepsilon_K$ for M8, M9 and M16. $M_{Z^\prime}=10\tev$ . Darker regions correspond to  present constraints on $\Delta M_{s,d}$ and lighter ones to future projection.
}\label{M8M9M1610Tev}~\\[-2mm]\hrule
\end{figure}

Now if the $B^0_{s,d}-\bar B^0_{s,d}$ mixing constraints dominate, which turns out still the case for $M_{Z^\prime}\le 30\tev$,
one has
\be
\tilde s_{13}^{\rm max}\propto M_{Z^\prime}, \qquad \tilde s_{23}^{\rm max}\propto M_{Z^\prime}\,, \qquad (\Delta M_{s,d}~~{\rm constraints}),
\ee
where we neglect RG effects. In this case
with increasing  $M_{Z^\prime}$: 
\begin{itemize}
\item
Maximal NP effects in $\epe$ increase slowly with RG effects represented by $r_{\varepsilon^\prime}$. 
\item
Maximal NP effects in $B_{s,d}$ decays decrease like $1/M_{Z^\prime}$ when their intereference with SM contribution dominates the modifications in the branching ratios.
\item
Maximal NP effects in $\kpn$ and $\klpn$ are independent of $M_{Z^\prime}$. 
\item 
Maximal NP effects in $\varepsilon_K$ and $\Delta M_K$ increase quadratically with  $M_{Z^\prime}$ up to the point at which $\tilde s_{13}$ and $\tilde s_{23}$ reach maximal values allowed by the 
unitarity of the new mixing matrix. But this point is never reached as in 
particular $\Delta M_K$ constraint becomes important with increasing $M_{Z^\prime}$ much earlier.
\end{itemize}

As maximal NP contributions to $\varepsilon_K$ and $\Delta M_K$ allowed by 
 $B^0_{s,d}-\bar B^0_{s,d}$ mixing constraints increase fast with increasing 
$M_{Z^\prime}$  these two observables will dominate the allowed ranges for  $\tilde s_{ij}$  at sufficiently high value of $M_{Z^\prime}$ and the pattern 
of  $M_{Z^\prime}$  dependences changes. Assuming for simplicity that the maximal
values of  $\tilde s_{13}$ and $\tilde s_{23}$ have the same  $M_{Z^\prime}$ 
dependence we have this time at fixed $\delta_2-\delta_1$
\be
\tilde s_{13}^{\rm max}\propto \sqrt{M_{Z^\prime}}, \qquad \tilde s_{23}^{\rm max}\propto \sqrt{M_{Z^\prime}}\,, \qquad (\varepsilon_K,\, \Delta M_K~~{\rm constraints}).
\ee
 In this case
with increasing  $M_{Z^\prime}$: 
\begin{itemize}
\item
Maximal NP effects in $\epe$ decrease up to  RG effects represented by $r_{\varepsilon^\prime}$ as $1/M_{Z^\prime}$.
\item
Maximal NP effects in $B_{s,d}$ decays decrease like $1/M^{1.5}_{Z^\prime}$ when the intereference with SM contribution dominates the modifications in the branching ratios.
\item
Maximal NP effects in $\kpn$ and $\klpn$ decrease as $1/M_{Z^\prime}$.  
\item 
Maximal NP effects in $\Delta M_{s,d}$ decrease as $1/M_{Z^\prime}$.
\end{itemize}

A closer inspection of formulae (\ref{eprimeZPlarge}), (\ref{triple}) and (\ref{triple2}) shows that it is the $\Delta M_K$ constraint that is most important. 
Indeed, in order to have NP in $\epe$ to be significant,  we need 
$\sin(\delta_2-\delta_1)\approx \pm 1$ with the sign  dependent on the model considered as 
listed in the last column in Table~\ref{tab:331models}. This is 
allowed by $B^0_{s,d}-\bar B^0_{s,d}$ mixing constraints. But then as seen in (\ref{triple}) the 
shift in $\varepsilon_K$ can be kept small in the presence of a significant shift in $\epe$ by having $\cos(\delta_2-\delta_1)$ very small. However, in 
the case of $\Delta M_K$ this is not possible as in this limit 
(\ref{triple2}) reduces to
\be\label{triple2a}
\frac{(\Delta M_K)^{Z^\prime}}{(\Delta M_K)_\text{exp}} =
\mp\frac{g^2(M_{Z^\prime}) c_W^2}{\beta}
\tilde s_{13} \tilde s_{23} \sin(\delta_2-\delta_1) \left[\frac{r_{\varepsilon}}{r_{\varepsilon^\prime}}\right]
\left[\frac{0.76}{\bei}\right] 1.49 \times 10^4\, \left(\frac{\varepsilon'}{\varepsilon}\right)_{Z^\prime}
\ee
with $-$ for $F_1$ and and $+$ for $F_2$. From the signs of $\beta$ and 
$\sin(\delta_2-\delta_1)$ in Table~\ref{tab:331models} 
 we find therefore that for large values of $M_{Z^\prime}$ significant enhancement of $\epe$ in 331 models implies uniquely suppression 
of $\Delta M_K$ for all 331 models considered and for sufficiently large values of  $M_{Z^\prime}$ this suppression will be too large to agree with data and this in turn will imply suppression  of $\epe$. This suppression of  $\Delta M_K$ is
not accidental and is valid in any $Z^\prime$ model in which flavour violating 
couplings of $Z^\prime$ to quarks are dominantly imaginary as one can easily 
derive from (\ref{Mk}). For $\sin(\delta_2-\delta_1)\approx \pm 1$ as required 
in 331 models to get large shift in $\epe$ the relevant couplings must 
indeed be dominantly imaginary but in general $Z^\prime$ models this could 
not necessarily be the case and also enhancements of $\Delta M_K$ could be
possible. 

When $\Delta M_{s,d}$, $\varepsilon_K$  and $\Delta M_K$ constraints are equally important  the pattern is more involved but these dependences indicate what we should 
roughly expect. The main message from this analysis is that with increasing $M_{Z^\prime}$ the importance of NP effects in $K$ meson system is likely to increase relative to the one in $B_{s,d}$ systems. But one should be cautioned that this depends also 
on other parameters and on the size of departures of SM predictions for various observables  from the data.

\begin{figure}[!tb]
 \centering
\includegraphics[width = 0.45\textwidth]{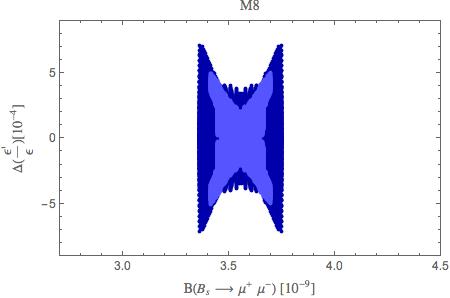}
\includegraphics[width = 0.45\textwidth]{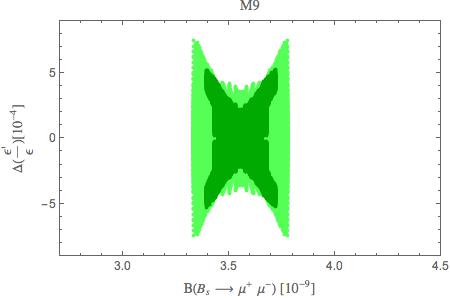}
\includegraphics[width = 0.45\textwidth]{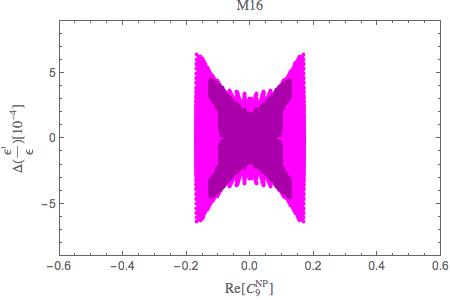}
\includegraphics[width = 0.45\textwidth]{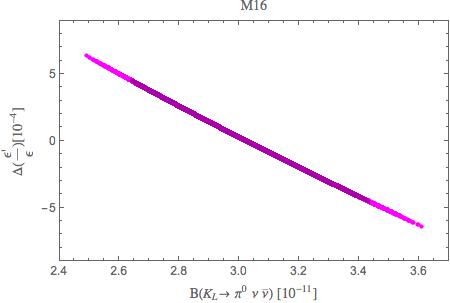}
\caption{ \it Correlations of $\Delta(\epe)$ with various observables  for M8, M9 and M16
at  $M_{Z^\prime}=10\tev$ .  Colour coding  as in Fig~\ref{M8M9M16}.
}\label{CORR10TEV}~\\[-2mm]\hrule
\end{figure}

Therefore, a detailed quantitative analysis of this pattern will only be possible when the room left for NP in the quantities in question will be better known.
But, the message is clear: possible tensions in $\epe$ and $\varepsilon_K$ 
can be removed in 331 models for values of $M_{Z^\prime}$ beyond the LHC   easier than in rare $B_{s,d}$ decays. 

As an example we show in Fig.~\ref{M8M9M1610Tev} $\Delta(\epe)$ versus $\varepsilon_K$ for the favourite models M8, M9 and M16 at $M_{Z^\prime}=10\tev$. We observe in accordance with our arguments that at this mass the maximal effects in $\epe$ found at $M_{Z^\prime}=3\tev$ are still possible and the range for possible 
values of $\varepsilon_K$ is significant increased. NP contributions to $\Delta M_K$ at these 
scales at  $M_{Z^\prime}=10\tev$ are still at most of $\pm 4\%$ and the 
$\Delta M_K$ constraint begins to play a role only for $M_{Z^\prime}\ge 30\tev$.

In Fig.~\ref{CORR10TEV} we show the correlation of $\Delta(\epe)$ with $B_s\to\mu^+\mu^-$ for $M_{Z^\prime}=10\tev$ in models M8 and M9 and the correlations 
of $\Delta(\epe)$ with $C_9$ and $\klpn$ in M16. To this end we imposed the 
constraints in (\ref{oases23}),  (\ref{oases13}) and  (\ref{CONE}).  
As expected from  $M_{Z^\prime}$ dependences discussed above, we observe that when the $B^0_{s,d}-\bar B^0_{s,d}$ mixing constraints 
still dominate, NP effects in $B_s\to\mu^+\mu^-$ and $C_9$ are significantly decreased while they remain practically unchanged in the case of $\epe$ and $\klpn$. 

With further increase of $M_{Z^\prime}$ NP effects in  $B_s\to\mu^+\mu^-$ and $C_9$ further decrease but the maximal effects in $\klpn$ are 
unchanged. In $\epe$ they even increase due to the increase of $r_{\varepsilon^\prime}$ so that for  $M_{Z^\prime}\approx 30\tev$ the shift in $\epe$ can 
reach approximately $8\times 10^{-4}$ in all three models.
But for higher values
of  $M_{Z^\prime}$ the $\Delta M_K$ constraint becomes important and NP effects in both 
$\epe$ and $\klpn$ are suppressed relative to the region  $M_{Z^\prime}\le 30\tev$ as expected from our discussion of the $M_{Z^\prime}$ dependence. We illustrate this in the case of M16 in Fig.~\ref{CORR50TEV} for  $M_{Z^\prime}=50\tev$. The 
suppression is slightly stronger in the case of $\klpn$ because in the 
case of $\epe$ it is compansated by roughly $10\%$ by the increase of 
$r_{\varepsilon^\prime}$.
The result for the first correlation in M8 and M9 is similar but NP effects in $\klpn$ are tiny in these 
models.

\begin{figure}[!tb]
 \centering
\includegraphics[width = 0.45\textwidth]{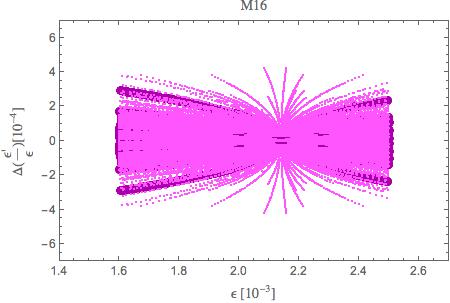}
\includegraphics[width = 0.45\textwidth]{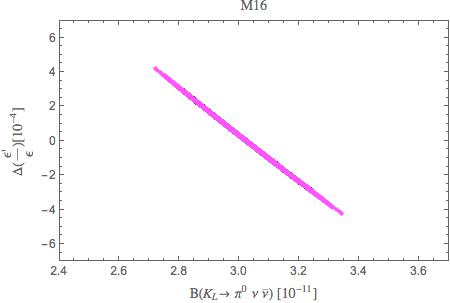}
\caption{ \it Correlations of $\Delta(\epe)$ with $\varepsilon_K$ and $\klpn$
in M16
at  $M_{Z^\prime}=50\tev$ .  Colour coding  as in Fig~\ref{M8M9M16}.
}\label{CORR50TEV}~\\[-2mm]\hrule
\end{figure}

\section{Summary}\label{sec:5}
In this paper we have updated and improved our analysis of $\epe$ in 331 models 
presented in \cite{Buras:2014yna}. The new analysis has been motivated by the 
new results for this ratio from \cite{Buras:2015yba,Blum:2015ywa,Bai:2015nea,Buras:2015xba} which show that $\epe$ within the SM is significantly below the 
data. 

Considering first seven 331 models selected by us in \cite{Buras:2014yna}
by electroweak precision tests and requiring a shift 
$\Delta (\epe)\ge 4.0\times 10^{-4}$ by NP, we reduced the number of 331 models to three: M8, M9 
and M16 in the terminology of \cite{Buras:2014yna}. All three can provide for 
$M_{Z^\prime}= 3\tev$ a shift in $\epe$ of $(5-6)\cdot 10^{-3}$ and this 
could in principle be sufficient to bring the theory to agree with data if
$\bsi$ increases towards its upper bound in the future. 
Moreover:
\begin{itemize}
\item
Models M8 and M9 can simultaneously suppress $B_s\to\mu^+\mu^-$ but do not offer 
the explanation of the suppression of the Wilson coefficient $C_9$ in $B\to K^*
\mu^+\mu^-$ (the so-called LHCb anomaly).
\item
On the contrary M16 offers an explanation of this anomaly simultaneously enhancing $\epe$ but does not provide suppression of  $B_s\to\mu^+\mu^-$ 
which could be required when the data improves and the inclusive value of
$\vcb$ will be favoured.
\item
NP effects in  $\kpn$,  $\klpn$ and $B\to K(K^*)\nu\bar\nu$ are small which 
can be regarded as prediction of these models to be confronted in the future
with NA62, KOPIO and Belle II results.
\item
Interestingly for values of $M_{Z^\prime}$  well above the LHC scales 
our favourite 331 models can still successfully 
face  the $\epe$ anomaly and also possible tensions in $\varepsilon_K$ 
can easier be removed than for  $M_{Z^\prime}$  in the reach of the LHC. 
This is clearly seen in Fig.~\ref{M8M9M1610Tev} obtained for $M_{Z^\prime}=10\tev$ and similar behaviour is found for  $M_{Z^\prime}$ up to $30\tev$  with maximal NP contribution to
 $\epe$ increased by RG effects up to $8\times 10^{-4}$. On the other hand, as seen in Fig.~\ref{CORR10TEV}, the effects in $B_s\to\mu^+\mu^-$ and $C_9$ are found above  $M_{Z^\prime}=10\tev$ to be very small. For 
 $M_{Z^\prime}=50\tev$ also effects in $\epe$ become too small to be able 
to explain the $\epe$ anomaly.
\end{itemize}

 The possibility of accessing masses of $M_{Z^\prime}$ far beyond the LHC reach in 
331 models with the help of $\epe$ and $\varepsilon_K$ is very appealing but one should keep in mind that the future 
of 331 models will crucially depend on the improved theory for $\epe$ and $\Delta F=2$ observables and improved data on rare $B_{s,d}$ and $K$ decays as we have stressed at various 
places of this writing, in particular when presenting numerous plots.

\section*{Acknowledgements}
We thank Christoph Bobeth for discussions.
This research was done and financed in the context of the ERC Advanced Grant project ``FLAVOUR''(267104) and was partially
supported by the DFG cluster
of excellence ``Origin and Structure of the Universe''.

\bibliographystyle{JHEP}
\bibliography{allrefs}
\end{document}